\documentclass{ws-procs975x65}

\usepackage{amssymb}
\usepackage{amsmath}
\usepackage{amsfonts}

\newcommand\beq{\begin{equation}}
\newcommand\eeq{\end{equation}}
\newcommand\bea{\begin{eqnarray}}
\newcommand\eea{\end{eqnarray}}
\newcommand\bseq{\begin{subequations}} 
\newcommand\eseq{\end{subequations}}
\newcommand\bal{\begin{align}}  
\newcommand\ealign{\end{align}}    

\begin{document}

\title{Pre-inflationary perturbations spectrum}
			
\author{GIOVANNI IMPONENTE}

\address{Dipartimento di Fisica, 
 						Universit\'a ``Federico II'', 
 						Napoli  and 
					 INFN -- Napoli -- Italy \\
					ICRA -- International Center for
								Relativistic Astrophysics \\
E-mail: imponente@icra.it}

\author{GIOVANNI MONTANI}

\address{Dipartimento di Fisica -- G9
					Universit\'a ``La Sapienza'', Roma -- Italy \\
 ICRA -- International Center for
								Relativistic Astrophysics \\
				E-mail: montani@icra.it}  


\maketitle

\abstracts{
In the framework of a flat FLRW model we derive 
an inflationary regime in which the scalar field, 
laying on the plateau of its potential, admits a linear 
time dependence and remains close to a constant 
value. \\
The behaviour of inhomogeneous perturbations is 
determined on the background metric in agreement 
to the ``slow-rolling''  approximation. 
We show that the inhomogeneous scales 
which before inflation were not much greater 
then the physical horizon, conserve their spectrum 
(almost) unaltered after the de Sitter phase. 
}

\section{Introduction}

The Standard Cosmological Model (SCM) finds many 
confirmations in the picture of the actual Universe
\cite{KT90}, but its shortcomings to describe very 
early stages of evolution appear as soon as the 
so-called  horizon and flatness paradoxes are taken 
into account \cite{G81,L82,L83}. \\
The Inflationary Paradigm (IP) has acquired progressively 
an increasing 
interest, because it provides a natural explanation 
for such paradoxes \cite{G81,L82,L83}; 
the IP success relies 
overall on the consistent and simultaneous treatment
of many different aspects of the cosmological puzzle. 
The capability to generate a
Harrison--Zeldovich Spectrum (HZS) 
for the density perturbations outstands among these,
and in fact the IP predicts it from the 
quantum fluctuations of the scalar field during 
the de Sitter phase. \\
This picture is well-grounded, but has to face 
the delicate point regarding the mechanism by which 
the quantum inhomogeneities approach a classical 
limit \cite{Sta79}.
The quantum origin of the perturbations is also 
supported by the exponential suppression of  
the ultra-relativistic inhomogeneities during 
the de Sitter phase \cite{IM03}. 

In this work we show the existence of an inflationary 
regime allowing a classical origin for the HZS. 
In fact we deal with perturbations of the scalar field 
$\phi$  which, if described by a HZS before inflation, 
survive to the Universe exponential 
expansion; in our solution the inhomogeneities 
become super-horizon--sized  and become seed for the 
structure formation when they re-enter
the horizon after the IP.
 
\section{Inflationary Regime} 
 
Let us consider a flat FLRW cosmology, summarized 
in a synchronous reference by the line element 
 \beq
 ds^2 = c^2 dt^2 - a^2(t) \delta_{ij}dx^i dx^j \, , \qquad i,j=1,2,3
 \eeq
where $c$ denotes the speed of light.\\
Let us introduce the adimensional scalar field
$\xi=\sqrt{\chi}\phi$ (being $\chi$ the Einstein 
constant) and analyse its dynamics on a plateau
region described by the potential
\beq
V(\xi) = \frac{3 \Lambda}{\chi c^2} - 
\frac{3 \lambda}{ \chi c^2} \frac{\xi^4}{4} \, ;
\eeq
this profile approximates a Coleman--Weimberg 
model and $\Lambda$ is the scale of the symmetry 
breaking, while $\lambda$ is a small coupling 
constant.\\
The dynamics of the coupled system $(a,\xi)$
is summarized by the field equations
\bseq
\bal
&\left( \frac{\dot{a}}{a}\right)^2 =
 \frac{\dot{\xi}}{6} +\Lambda - \frac{\lambda}{4}\xi^4 \\
 &\ddot{\xi} + 3 \frac{\dot{a}}{a}\dot{\xi}
 - 3 \lambda \xi^3 = 0 \, ,
\end{align}
\label{feq}
\eseq
where $\dot{(~)}\equiv d/dt$.\\
The solutions of (\ref{feq}) read as 
\bseq
\bal
a(t) &= a_0 e^{H(t-t_0)} \, , \qquad a_0, t_0= \textrm{const.}\\
\xi(t) &=\alpha + \frac{\lambda}{H}\alpha^3 (t-t_0)
\end{align}
\eseq
with $\alpha=\textrm{const.}$ 
$(\alpha \ll \sqrt[4]{\Lambda/\lambda})$ and 
\beq
H= \frac{1}{\sqrt{2}}\left[ \left( \Lambda - \frac{\lambda}{4} \alpha^4\right)
+ \sqrt{ \left( \Lambda - \frac{\lambda}{4} \alpha^4\right)^2+
\frac{2}{3} \lambda^2 \alpha^6} 
\right]^{1/2} \sim \sqrt{\Lambda} \, .
\eeq
The obtained de Sitter evolution holds
as far as 
\beq
t-t_0  \ll \frac{H}{\lambda \alpha^2} \sim 
\frac{\sqrt{\Lambda}}{\lambda \alpha^2}
\eeq
and  an e-folding of order 
$\mathcal{O}(10^2)$ implies that 
\beq
\alpha \ll \mathcal{O} \left(\frac{1}{10}
\sqrt{\frac{\Lambda}{\lambda}} \right) \, , 
\eeq
already ensured by the structure of the solution.\\
Here we got a consistent inflationary dynamics
during which the scalar field remains close to 
a constant value $\phi \sim \alpha/\sqrt{\chi}$.\\
When $t-t_0$ increases enough, the scalar field
enters a different regime of the IP, which ends 
with the fall-down into the true vacuum an the 
associated re-heating process. 

\section{Perturbations Dynamics}

Now we study the behaviour of inhomogeneous 
perturbations of the scalar field, by 
neglecting their back-reaction on the scale 
factor $a(t)$.\\
If we take $\xi \rightarrow \xi + \delta(t,x^i)$, 
then this perturbation has to satisfy the 
dynamics
\beq
\ddot{\delta}+ 3H \dot{\delta} -
\left( \frac{c^2 }{a^2}\delta^{ij} \partial_i
\partial_j + 
9 \lambda \alpha^2
\right) \delta =0 \,;
\label{dde}
\eeq
the Fourier transform of (\ref{dde})
satisfies the equation
\beq
\ddot{\Delta}+ 3H \dot{\Delta} + 
\left( \frac{\mid \vec{k}\mid ^2}{a^2}c^2 - 
9 \lambda \alpha^2
\right) \Delta =0
\label{ddd}
\eeq
where $\Delta (t,\vec{k})$ is Fourier conjugated to 
$\delta(t,x^i)$. Equation (\ref{ddd}) admits the 
(approximate) solution 
(with $\ddot{\Delta}\sim 0$)
\beq
\Delta (t, \vec{k}) =\varepsilon (\vec{k})~
\textrm{exp}\left[ \frac{\mid \vec{k}\mid ^2 c^2}{
6 H^2 a_0^2} \left( e^{-2H(t-t_0)}-1\right) +
3 \frac{\lambda}{H}\alpha^2(t-t_0)
\right]
\eeq
where $ \varepsilon (\vec{k}) \equiv \Delta(t_0,\vec{k})$
denotes the pre-inflationary spectrum.\\
At the end of the de Sitter phase, we get 
the following (to leading order) spectrum 
$\Delta_f(\vec{k})$ in terms of the initial one
$\Delta_i(\vec{k})$
\beq
\Delta_f(\vec{k}) \sim \Delta_i(\vec{k})
\textrm{exp}\left(-{\frac{k^2 }{6 H^2 a_0^2}}\right) \, ;
\eeq
all the physical scales 
$\lambda_{ph}\equiv \frac{2 \pi}{\mid \vec{k}\mid}a_0$,
which where not much greater than the physical 
horizon $cH^{-1}$ at the beginning of the IP,
survive with (almost) their spectrum.\\
If we deal with a HZS, the above analysis shows 
that its classical origin is compatible with 
an IP.

\end{document}